\documentclass[prb,reprint,superscriptaddress,twocolumn,nofootinbib]{revtex4-2}
\usepackage{amsmath,amssymb,bm,braket,mathrsfs,mathtools}
\usepackage{multirow}
\usepackage{color}
\usepackage{url}
\usepackage{hyperref}
\usepackage{amsmath,bbm,mathrsfs}
\hypersetup{
	setpagesize=false,
	bookmarksnumbered=true,%
	bookmarksopen=true,%
	colorlinks=true,%
	linkcolor=magenta,
	citecolor=magenta,
}

\DeclareMathOperator{\tr}{tr}

\DeclareMathOperator{\pf}{Pf}
\allowdisplaybreaks

\begin{document}
\title{Euler band topology in superfluids and superconductors}
\author {Shingo Kobayashi}
\affiliation{RIKEN Center for Emergent Matter Science, Wako, Saitama 351-0198, Japan}
\author{Manabu Sato} 
\affiliation{Department of Applied Physics, University of Tokyo, Bunkyo, Tokyo 113-8656, Japan}
\author {Akira Furusaki}
\affiliation{RIKEN Center for Emergent Matter Science, Wako, Saitama 351-0198, Japan}

\date{\today}
\begin{abstract} 
Real band topology often appears in systems with space-time inversion symmetry and is characterized by invariants such as the Euler and second Stiefel–Whitney classes.
Here, we examine the generic band topology of Bogoliubov de-Gennes (BdG) Hamiltonians with $C_{2z}T$ symmetry, where $C_{2z}$ and $T$ are twofold rotation about the $z$ axis and time-reversal symmetries, respectively.
We discuss the Euler band topology of superfluids and superconductors in the DIII and CI Altland-Zirnbauer symmetry classes,
where the Euler class serves as an integer-valued topological invariant of the $4\times4$ BdG Hamiltonian.
Using expressions for the Euler class under $n$-fold rotational symmetry, we derive formulas relating the Euler class to previously known topological invariants of class DIII and CI systems.
We demonstrate that three-dimensional class DIII topological phases with an odd winding number, including the B phase of superfluid Helium 3, are topological superconductors or superfluids with a nontrivial Euler class.  We refer to these as Euler superconductors or superfluids.
Specifically, the $^3$He-B superfluid in a magnetic field is identified as an Euler superfluid.
Three-dimensional class CI topological phases with twice an odd winding number are also Euler superconductors or superfluids.
When spatial inversion symmetry is present, class CI superconductors with a nontrivial Euler class exhibit superconducting nodal lines with a linking structure. This phenomenon is demonstrated using a model of a three-dimensional $s_\pm$-wave superconductor.
These findings provide a unified framework for exploring Euler band topology in superfluids and superconductors, connecting various phenomena associated with $T$-breaking perturbations, including Majorana Ising susceptibility and higher-order topology.
\end{abstract}
\maketitle
\textit{Introduction.}
Real band topology arises in electron systems with space-time inversion symmetry $I_{\rm ST}$ that squares to identity operator.
In such systems energy bands form real vector bundles, whose topology is characterized by
the Euler class for $N=2$ and 
the second Stiefel-Whitney (SW) class for $N>2$, where $N$ is the number of occupied bands~\cite{Hatcher2002,Morimoto2014prb,CFang2015,Fang2015new,Zhao2017pt,Bzdusek2017robust,Ahn2018band,Bouhon2019,Ahn2019failure,Ahn2019stiefel,Bouhon2020}. A pair of real bands with a nonzero Euler class, referred to as Euler bands, exhibits distinctive topological phenomena such as violation of the fermion doubling theorem and topological phase transitions through braiding of band degeneracy points~\cite{Wu2019non-abelian,Ahn2019failure,bouhon2020non}. 

Over the past several years, the discovery of Euler band topology in twisted bilayer graphene~\cite{Po2019,Ahn2019failure,Song2019all} has generated significant interest in $I_{\rm ST}$-symmetry protected topological phases across a wide range of systems. These include not only electronic bands in topological insulators~\cite{Ahn2019,Kobayashi2021fragile,Brouwer2023,sato2025euler}, topological semimetals~\cite{Wu2019non-abelian,bouhon2020non,Chen2022non-abelian,sato2024ideal,ghadimi2024,mondal2024}, and magnetic materials~\cite{Lee2025Euler}, but also phonon bands in elastic crystals~\cite{park2021topological} and layered materials~\cite{peng2022phonons,Peng2022Multigap}.
Other systems of relevance are artificial platforms, such as cold atomic setups~\cite{Unal2020,karle2024anomalous}, acoustic metamaterials~\cite{Valerio2020Experimental,jiang2021experimental,Bin2024Observation,qiu2023minimal,Davoyan2024}, photonic crystals~\cite{park2021non,Wang2022Experimental,xu2023absence,hu2024observation,Yi2024Non-Abelian,Liu2025Correspondence}, transmission line networks~\cite{guo2021experimental,jiang2021four}, electric circuits~\cite{Ezawa2021}, and spring-mass systems~\cite{Park2022topological}.

Implications of the Euler band topology in superconductors have also been explored in several recent studies.
The presence of superconducting gap nodes dictated by real band topology was
identified through homotopy theory~\cite{Bzdusek2017robust}.
Subsequent studies, motivated by the developments in twisted bilayer graphene, have discussed obstructions to Cooper pairing%
~\cite{Lo2022Inherited,Yu2023Euler-obstructed} 
and nodal higher-order topological superconductivity in Euler bands~\cite{Yu2022Euler-obstructed}.
More recently, 
Andreev reflection~\cite{Morris2024Andreev} and non-linear optical responses~\cite{chau2025optical} of superconducting Euler bands have been discussed.
However, most of these studies are based on specific models and assume nontrivial Euler topology in the normal-state Hamiltonian. 
It remains an open question how Euler band topology emerges in a more general setting in superconductors and superfluids.

\begin{figure}[tbp]
  \begin{center}
    \includegraphics[width=\linewidth]{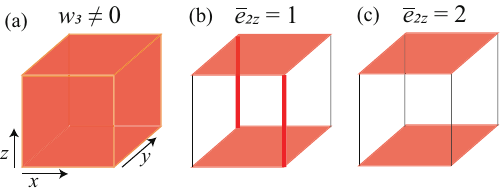}
    \caption{
    (Color online) 
    (a) Schematic illustration of gapless surface states (red) in topological phases with a nonzero 3D winding number [Eq.~(\ref{eq:w3d})]. (b) and (c) Gapless surface states under application of a $T$-breaking but $C_{2z}T$-preserving perturbation [e.g., magnetic field in the [110] direction] to the Euler superconductor (superfluid) shown in (a), as implied by the relation between the 3D winding number and Euler class in Eqs.~(\ref{eq:Euler-DIII_3D}) and (\ref{eq:Euler-CI_3D}).
We infer the existence of hinge states from the relation between the second SW class and the Chern-Simons invariant~\cite{Ahn2019}, where $\bar{e}_{2z}$ is an odd integer.
    }
    \label{fig:surface}
  \end{center}
\end{figure}

In this letter, we discuss the real band topology of the Bogoliubov-de Gennes (BdG) Hamiltonians with $I_{\rm ST}$ symmetry, focusing on time-reversal-symmetric superconductors in classes DIII and CI in the Altland-Zirnbauer classification~\cite{Altland97}, for which an $I_{\rm ST}$-operator is defined with additional two-fold rotation or spatial inversion symmetry.
In most part of our discussions we take $I_{\rm ST}$ to be the product of two-fold rotation and time-reversal operations,
which is strictly speaking $I_{\rm ST}$ in two spatial dimensions.
The Euler class is defined by the integral of real Berry curvature over an $I_{\rm ST}$-invariant two-dimensional (2D) plane in the three-dimensional (3D) Brillouin zone (BZ).
We derive simplified formulas for the Euler class for BdG Hamiltonians under $I_{\rm ST}$ and $n$-fold rotation symmetries ($n=2,4$).
These formulas establish explicit relationships between the Euler class and stable topological invariants in class DIII.
Notably, we show that the 3D winding number is equivalent to the Euler class modulo two, indicating the persistence of surface states under $I_{\rm ST}$-symmetry-preserving, but time-reversal-symmetry-breaking (TRSB) perturbations [Figs.~\ref{fig:surface} (a) and (b)]. As a concrete example, we show that the B phase of $^3$He possesses a nontrivial Euler class.

We extend our analysis to class CI systems with representation-protected topology~\cite{kobayashi2024}. 
The Euler class is related to the 3D winding number characterizing 3D topological phases [Fig.~\ref{fig:surface} (c)]. 
When spatial inversion symmetry is present, the product of the time-reversal and spatial inversion operations plays a role of the space-time inversion $I_{\rm ST}$, in place of time-reversal and two-fold rotation operations.
In this case the full 3D BZ is an $I_{\rm ST}$-invariant space.
A nonzero Euler class on a 2D closed sphere in the 3D BZ is shown to indicate that this superconductor has a nodal line linking with degeneracy lines of energy bands, inside the 2D sphere.

\textit{Euler class for BdG Hamiltonians.}
We begin by outlining the symmetries relevant to our analysis. The fundamental symmetries of superfluids and superconductors are the particle-hole ($C$) and time-reversal ($T$) symmetries. 
Both $C$ and $T$ are antiunitary operators and satisfy the relations $[T,C]=0$, $C^2=(-1)^{s+1}$ and $T^2=(-1)^{s}$ for class CI ($s=0$) and DIII ($s=1$), where $s=0\, (1)$ indicates the presence (absence) of SU(2) spin-rotation symmetry.
These symmetry operations act on the BdG Hamiltonian in the 3D momentum space, $H_{\text{BdG}} = \frac{1}{2}\sum_{\bm{k},\alpha,\beta} \Psi^{\dagger}_{\bm{k},\alpha} \mathcal{H}_{\alpha \beta}(\bm k) \Psi_{\bm{k},\beta}$, where 
\begin{align}
\mathcal{H}_{\alpha \beta}(\bm k) &=  \begin{pmatrix} 
  \mathcal{E}_{\alpha \beta}(\bm k) & \Delta_{\alpha \beta}(\bm k)  \\
  \Delta_{\alpha \beta}^{\dagger}(\bm k) & - [\mathcal{E}^{ t}(-\bm k)]_{\alpha \beta} 
\end{pmatrix},
\label{eq:BdG}
\end{align}
and $\Psi_{\bm{k},\alpha}^t = (c_{\bm{k},\alpha},c^{\dagger}_{-\bm{k},\alpha})$. The superscript $t$ denotes transposition, and the indices $\alpha, \beta$ represent spin, orbital, and sublattice degrees of freedom. The normal-state Hamiltonian $\mathcal{E}(\bm k )$ and the pair potential $ \Delta({\bm k})$ satisfy $\mathcal{E}(\bm k ) = \mathcal{E}^{\dagger}(\bm k )$ and $\Delta^{t}(\bm k) =(-1)^{s}\Delta(-\bm k)$ due to the Fermi-Dirac statistics.
Note that in class CI, the Hamiltonian is block-diagonalized into two spin sectors, in each of which $\Delta^{t}(\bm k) =\Delta(-\bm k)$. 
In this basis, $C$ and $T$ are represented as
\begin{align}
    &C \mathcal{H}(\bm{k}) C^{-1} = \mathcal{H}(-\bm{k}) ,  \ \ C= \begin{cases} i\tau_y K &\text{ if } s = 0, \\   \tau_x  K & \text{ if } s = 1, \end{cases} \label{eq:phs}\\
    &T \mathcal{H}(\bm{k}) T^{-1} = \mathcal{H}(-\bm{k}) , \ \ T= \begin{pmatrix} U_T & 0 \\ 0 & U_T^{\ast}\end{pmatrix} K , \label{eq:trs}
\end{align}
where $\tau_i$ are Pauli matrices in the Nambu space, $K$ denotes complex conjugation, $U_T$ is a unitary matrix satisfying $U_T U_T^{\ast} = (-1)^{s} \mathbbm{1}_N$, and $\mathbbm{1}_N$ is the $N \times N$ identity matrix, with $N$ being the matrix dimension of $\mathcal{E}(\bm{k})$. 

In addition to these fundamental symmetries, we consider space-time inversion symmetry $I_{\rm ST}$.
This is an antiunitary operator satisfying the relations 
\begin{align}
  I_{\rm ST}^2=+1 \ \ \text{and} \ \ [I_\mathrm{ST},C]=[I_\mathrm{ST},T]=0.
   \label{eq:space-time}
\end{align}
For class CI, we take $I_{\rm ST}$ to be either $C_{2z}T$ or $PT$, where $C_{2 z}$ denotes twofold rotation about the $z$ axis and $P$ is spatial inversion.
These operators satisfy $(C_{2z})^2= (-1)^{s}$ and $P^2 = 1$, thereby $I_\mathrm{ST}^2=+1$.
For class DIII, we can only take $I_\mathrm{ST}=C_{2z}T$, because $(PT)^2=-1$.
The matrix representation of $I_{\rm ST}$ is obtained by combining Eq.~(\ref{eq:trs}) with 
\begin{align}
    \mathcal{G} \mathcal{H}(\bm{k}) \mathcal{G}^{-1} = \mathcal{H}(R_g \bm{k}), \ \ \mathcal{G} = \begin{pmatrix} U_g & 0 \\ 0 & U_g^{\ast}\end{pmatrix}, \label{eq:inv-rot}
\end{align} 
where $g=C_{2z}$ or $P$, and $U_g$ is a unitary matrix satisfying $U_{C_{2z}}^2= (-1)^{s} \mathbbm{1}_N$ and $U_P^{2}=\mathbbm{1}_N$.
The matrix $R_g\in\mathrm{O}(3)$ acts on $\bm{k}$ as $R_{C_{2z}} \bm{k} = (-k_x,-k_y,k_z)$ and $R_P \bm{k} = - \bm{k}$.
Note that $U_g U_T$ is symmetric and $U_g U_T=U_T U_g^*$, which follows from $(gT)^2=+1$ and $[g,T]=0$, respectively.

We now consider the band topology of the BdG Hamiltonian in the presence of $I_{\rm ST}$ symmetry. In the basis where the chiral operator, defined as the product of the $\mathcal{C}$ and $\mathcal{T}$ operators, is diagonal, the Hamiltonian takes the off-diagonal form
\begin{align}
    &
    \mathcal{H}(\bm{k})= V_{\Gamma}^{\dagger} 
     \begin{pmatrix} 0 & A(\bm{k}) \\ A^{\dagger}(\bm{k}) & 0 \end{pmatrix} V_{\Gamma}, 
     \label{eq:off-diagnal} \\
    &
    V_{\Gamma} = \frac{1}{\sqrt{2}}
     \begin{pmatrix}
       \mathbbm{1}_N & - i U_T^t \\ \mathbbm{1}_N & i U_T^t
     \end{pmatrix}
\end{align}
with $A(\bm{k}) \equiv \mathcal{E}(\bm{k})-i U_T^t \Delta^{\dagger}(\bm{k})$.
We apply singular value decomposition to $A(\bm{k})$ and write it in the form $A(\bm{k}) = U^\dagger(\bm{k})D(\bm{k})V (\bm{k})$, where $U(\bm{k}),V(\bm{k}) \in \mathrm{U}(N)$ and $D(\bm{k})$ is a non-negative diagonal matrix. For fully gapped systems, we can make continuous deformation $D(\bm{k}) \to \mathbbm{1}_N$ without closing the energy gap. Then, $A(\bm{k})$ becomes a unitary matrix $Q(\bm{k}) = U^{\dagger}(\bm{k})V (\bm{k}) \in \mathrm{U}(N)$. In the $I_{\rm ST}$-invariant subspace of the 3D BZ, $Q$ is a symmetric unitary matrix satisfying
\begin{align}
    Q(\bm{k}_{\rm ST}) U_g U_T = [Q(\bm{k}_{\rm ST}) U_g U_T]^t
\end{align}
with $\bm{k}_{\rm ST}$ denoting $I_{\rm ST}$-invariant momenta.
Thus, it can be decomposed as
\begin{align}
    Q(\bm{k}_{\rm ST}) U_g U_T = \mathcal{U}(\bm{k}_{\rm ST})\mathcal{U}^t(\bm{k}_{\rm ST}) \label{eq:decomp}
\end{align}
with $\mathcal{U}(\bm{k}_{\rm ST}) \in \mathrm{U}(N)$. Since Eq.~(\ref{eq:decomp}) is invariant under $\mathcal{U} \to \mathcal{U O}$ for any $\mathcal{O} \in \mathrm{O}(N)$, we conclude that $Q(\bm{k}_{\rm ST}) U_g U_T \in \mathrm{U}(N)/\mathrm{O}(N)$.
Therefore, the topology of the gapped BdG Hamiltonians can be found from the homotopy of $\mathrm{U}(N)/\mathrm{O}(N)$.
In particular, its second homotopy group is given by \cite{Hatcher2002,Bzdusek2017robust}
\begin{align}
    \pi_2 \biglb(\mathrm{U}(N)/\mathrm{O}(N)\bigrb) = \pi_1 \biglb(\mathrm{O}(N)\bigrb) =
    \begin{cases} \mathbb{Z} & \text{if }  N=2, \\ \mathbb{Z}_2 & \text{if }  N \ge 3, \end{cases}
\end{align}
encoding the topology of BdG Hamiltonians on 2D $I_\mathrm{ST}$-invariant subspaces.
Here we have used the exact sequence of the homotopy groups.  
The $\mathbb{Z}$-valued invariant is called the Euler class, while the $\mathbb{Z}_2$-valued one is the second SW class.

Let us focus on the case $N=2$ and discuss the Euler class in detail by extending the framework developed for the Euler insulators~\cite{Ahn2019failure} to superconductors.
After the deformation from $A$ to $Q$, the eigenstates of Eq.~(\ref{eq:off-diagnal}) with an eigenvalue $-1$ are given by~\cite{Dai2021Takagi}
\begin{align}
   | n,\bm{k}_{\rm ST} \rangle = \frac{i}{\sqrt{2}} 
    V_{\Gamma}^{\dagger}  \left(\begin{array}{@{\,}c @{\,}} 
         -\mathcal{U}(\bm{k}_{\rm ST}) \varphi_n \\
         U_g U_T \mathcal{U}^{\ast}(\bm{k}_{\rm ST}) \varphi_n 
    \end{array}\right)
    \quad (n=1,2),
    \label{eq:real-gauge}
\end{align}
where $\varphi_n$ are two-component unit vectors, $\varphi_1=(1,0)^t$ and $\varphi_2=(0,1)^t$.
It is important to note that the real-gauge condition is imposed,
$\mathcal{G}T  | n,\bm{k}_{\rm ST} \rangle =| n,\bm{k}_{\rm ST} \rangle $.
The Euler class $e_2$ is then given by the flux integral over
an $I_{\rm ST}$-invariant 2D plane (e.g., $k_z=0$ or $\pi$ plane) in the 3D BZ,
\begin{align}
    e_2 = \frac{1}{2\pi} \int_{\rm 2Dinv} d \bm{S} \cdot \bm{F}_{12}(\bm{k}_{\rm ST} ), \label{eq:Euler_class}
\end{align}
where the real Berry curvature
$\bm{F}_{mn} \equiv \bm{\nabla}_{\bm{k}} \times \bm{A}_{mn}$
and the real Berry connection
$\bm{A}_{mn} \equiv\langle m,\bm{k}_{\rm ST} |\bm{\nabla}_{\bm{k}}| n,\bm{k}_{\rm ST} \rangle $.
Note that the Euler class is well-defined only for orientable real states, and the systems examined below satisfy the orientability condition \cite{suppl}, where we have used the arguments from Refs.~\cite{Ahn2018band} and \cite{Sato2010}.

\begin{figure}[tbp]
  \begin{center}
    \includegraphics[width=\linewidth]{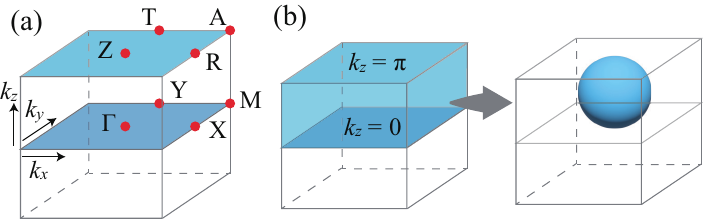}
    \caption{
    (Color online) (a) TRIMs in the 3D BZ of orthorhombic systems. (b) Deformation of the $k_z=0,\pi$ planes to a sphere.
    }
    \label{fig:trim}
  \end{center}
\end{figure}    

Equation (\ref{eq:Euler_class}) is reduced to a simpler form in rotation symmetric systems~\cite{Ahn2019failure,sato2025euler}, where an $n$-fold rotation operation is denoted as $g=C_{nz}$ in Eq.~(\ref{eq:inv-rot}), with the condition $(U_{C_{nz}})^n = (-1)^{s} \mathbbm{1}_N$.

For $C_{2z}$-symmetric class DIII superconductors ($s = 1$), we apply the Stokes' theorem to the integral in Eq.~(\ref{eq:Euler_class}) 
over a half of the $I_\mathrm{ST}$-invariant 2D plane. The resulting expression can be written in terms of $C_{2z}$ eigenvalues or, equivalently, the Pfaffian of the sewing matrix, yielding the formula
\begin{align}
 (-1)^{e_2} = \prod_{\bm{k}_{\rm T} \in \text{2Dinv}} \pf \left[ \mathcal{B}_{C_{2z}}(\bm{k}_{\rm T}) \right],
 \label{eq:c2z_e2}
\end{align}
where the sewing matrix is defined by
\begin{align}
   [\mathcal{B}_g(\bm{k}_{\rm ST})]_{mn} &= \langle m,R_g\bm{k}_{\rm ST} |  \mathcal{G} | n,\bm{k}_{\rm ST} \rangle \nonumber \\
    &=\left\{\text{Re}\left[\mathcal{U}^{\dagger}(R_g\bm{k}_{\rm ST})U_g \mathcal{U}(\bm{k}_{\rm ST})\right]\right\}_{mn}. \label{eq:sewing_matrix}
\end{align}
In the $I_{\rm ST}$-invariant subspace, $\mathcal{B}_{C_{2z}}(\bm{k}_{\rm ST})$ is a real orthogonal (due to the real-gauge condition) and skew-symmetric matrix (because $C_{2z}^2=-1$).
In Eq.\ (\ref{eq:c2z_e2}) $\bm{k}_{\rm T}$ denotes time-reversal-invariant momenta (TRIMs); for instance, $\bm{k}_{\rm T} \in \{\Gamma, \mathrm{X, Y, M}\}$ in the $k_z=0$ plane in Fig.~\ref{fig:trim}(a).

Similarly, for $C_{4z}$-symmetric class CI superconductors ($s = 0$), Eq.\ (\ref{eq:Euler_class}) can be written, after reducing the integral over the $I_\mathrm{ST}$-invariant $k_z=0$ plane to its quarter using the $C_{4z}$ symmetry, in the form
\begin{align}
    \cos \left( \frac{\pi}{2} e_2 \right)
     = -\xi_{C_{2z}}(\mathrm{X}) \pf \left[ \mathcal{B}_{C_{4z}}(\Gamma) \right]
       \pf \left[ \mathcal{B}_{C_{4z}}(\mathrm{M}) \right], \label{eq:c4z_e2}
\end{align}
where $\xi_{C_{2z}} (\mathrm{X}) \in \{\pm1\}$ is a two-fold rotation eigenvalue at the X point.
In deriving Eq.~(\ref{eq:c4z_e2}) we have assumed the condition $(U_{C_{4z}})^2 = -\mathbbm{1}_2$, which holds if the relevant two energy bands are formed by the orbitals from a specific two-dimensional irreducible representation of $C_{4z}$.
A more detailed explanation of this condition will be provided later. 
Below, we examine the connection between the Euler class and other known topological invariants in classes DIII and CI,
and discuss its physical implications.

\textit{Topology of class DIII systems.}
 We first focus on class DIII systems. This symmetry class has topological phases in one, two, and three spatial dimensions, which are characterized by $\mathbb{Z}_2$, $\mathbb{Z}_2$, and $\mathbb{Z}$, respectively~\cite{Schnyder08,Kitaev09,Schnyder09,Ryu10}. Examples include spin-triplet superconductors with helical pairing~\cite{Schnyder08,Qi2009}, the superfluid $^3$He-B phase~\cite{Volovik03,Schnyder08,Mizushima16}, and superconducting topological materials~\cite{sasaki2015sc,yonezawa2016bulk,SatoFujimoto16,Sato17}. 

First, we briefly review the topological invariants of class DIII. Using the unitary matrix $Q(\bm{k})$ and the skew-symmetry of $Q(\bm{k}_{\rm T})U_T$, the 1D $\mathbb{Z}_2$ invariant is defined by~\cite{Qi2010}
\begin{align}
    (-1)^{\nu_{12}} =&
     \frac{\pf[Q(\bm{k}_{\rm T1})U_T] }{\pf[Q(\bm{k}_{\rm T2})U_T]}
     \exp \!\left( \int_{\bm{k}_{\rm T1}}^{\bm{k}_{\rm T2}} \frac{dk}{2} \partial_k \ln \{ \det [Q(\bm{k})]\} \!\right)\!,
     \label{eq:DIII_1D}
\end{align}
where $\bm{k}_{\rm T1}, \bm{k}_{\rm T2}$ are TRIMs in the 1D BZ.
Equation~(\ref{eq:DIII_1D}) is extended to 2D BZs, where there are four TRIMs, labeled as $\bm{k}_{\rm T1-4}$. The 2D $\mathbb{Z}_2$ invariant is given by
\begin{align}
    (-1)^{\nu_{\rm 2d}} = (-1)^{\nu_{12} + \nu_{34} }. \label{eq:DIII_2D}
\end{align}
The 3D winding number $w_{\rm 3d}$ defined by
\begin{align}
    w_{\rm 3d} = &\int \frac{d^3k}{24 \pi^2} \sum_{\alpha, \beta, \gamma} \epsilon^{\alpha \beta \gamma} \tr\Big\{
    \left[Q^{\dagger}(\bm{k}) \partial_{k_\alpha} Q(\bm{k})\right] \nonumber \\
    &\times \left[Q^{\dagger}(\bm{k}) \partial_{k_\beta} Q(\bm{k})\right] \!
    \left[Q^{\dagger}(\bm{k}) \partial_{k_\gamma} Q(\bm{k})\right] \Big\} \label{eq:w3d}
\end{align}
takes integer values,
where $\alpha, \beta, \gamma=x,y,z$ and $\epsilon^{\alpha \beta \gamma}$ is the Levi-Civita symbol.
On the other hand, 
given the eight TRIMs ($\bm{k}_{\rm T1},\ldots,\bm{k}_{\rm T8}$) in the 3D BZ, we can define a 3D $\mathbb{Z}_2$ invariant as $\nu_{\rm 3d} = \nu_{12}+\nu_{34}+\nu_{56}+\nu_{78} \mod 2$. This invariant is related to the parity of the 3D winding number by~\cite{Qi2010}
\begin{align}
    (-1)^{w_{\rm 3d}} = (-1)^{\nu_{\rm 3d}}. \label{eq:DIII_3D}
\end{align}

We now prove a relation between these topological invariants and the Euler class in $C_{2z}$ symmetric systems, in which 
there are two $C_{2z}T$ invariant planes: $k_z=0$ and $\pi$ as shown in Fig.~\ref{fig:trim} (a).
Using Eq.~(\ref{eq:decomp}), the integrand in Eq.~(\ref{eq:DIII_1D}) is recast into
\begin{align}
     \frac{1}{2} \partial_k \ln \{ \det [Q(\bm{k})]\}  &= \frac{1}{2} \partial_k \ln \{ \det [\mathcal{U}(\bm{k})\mathcal{U}^t(\bm{k})   U_{ T}^{\dagger}U_{C_{2z}}^{\dagger}]\} \nonumber  \\
     &= \partial_k \ln \{ \det [\mathcal{U}(\bm{k})]\}, \label{eq:change}
\end{align}
where $U_{C_{2z}} U_{\rm T}$ is momentum-independent\footnote{In general, the symmetry operator depends on momentum, i.e., $U_{C_{2z}} (\bm{k})$. In this case, Eq. (\ref{eq:change}) has another contribution, $\frac{1}{2}\partial_k \ln \{\det [U^{\dagger}_{C_{2z}} (\bm{k})]\}$, which, however, does not affect the final result since $\exp\left(\int_{\bm{k}_{\rm T1}}^{\bm{k}_{\rm T2}} \frac{dk}{2} \, \partial_k \ln \{\det [U^{\dagger}_{C_{2z}} (\bm{k})]\} \right) =
\{\det [U^{\dagger}_{C_{2z}} (\bm{k}_{\rm T1})]/ \det [U^{\dagger}_{C_{2z}} (\bm{k}_{\rm T2})] \}^{1/2} =1$ due to the orientability of the real state.}.
Substituting Eq.\ (\ref{eq:change}) into Eq.~(\ref{eq:DIII_1D}), we obtain
\begin{align}
    (-1)^{\nu_{12}} 
    &=\frac{\pf[\mathcal{U}^{\dagger}(\bm{k}_{\rm T1})Q(\bm{k}_{\rm T1})U_T\mathcal{U}^\ast(\bm{k}_{\rm T1})] }{\pf[\mathcal{U}^{\dagger}(\bm{k}_{\rm T2})Q(\bm{k}_{\rm T2})U_T\mathcal{U}^\ast(\bm{k}_{\rm T2})]} \nonumber \\
     &= \frac{\pf [\mathcal{B}_{C_{2z}} (\bm{k}_{\rm T1})]}{\pf[\mathcal{B}_{C_{2z}} (\bm{k}_{\rm T2})]}, \label{eq:DIII_1Dv2}
\end{align}
where we have used the identities $\det[\mathcal{U}]^{-1} = \det[\mathcal{U}^{\dagger}]$ and $\det[B]\pf[A] = \pf[BAB^t]$ in the first line and Eqs.~(\ref{eq:decomp}) and (\ref{eq:sewing_matrix}) in the second line, noting that $\mathcal{U}^{\dagger}(\bm{k}_{\rm T}) U_{C_{2z}} \mathcal{U}(\bm{k}_{\rm T})$ is real due to $U_{C_{2z}}^2 = -\mathbbm{1}_2$.
When $N=2$, we compare Eq.~(\ref{eq:c2z_e2}) with Eqs.~(\ref{eq:DIII_2D}), (\ref{eq:DIII_3D}), and (\ref{eq:DIII_1Dv2}) to find that the Euler class satisfies
\begin{align}
    (-1)^{e_2(\bar{k}_z)} = (-1)^{\nu_{2d}(\bar{k}_z)},  \label{eq:Euler-DIII_2D}
\end{align}
in the 2D $C_{2z}T$ invariant planes ($\bar{k}_z=0$ or $\pi$), and
\begin{align}
     (-1)^{\bar{e}_{2z}}  = (-1)^{w_{3d}},  \label{eq:Euler-DIII_3D}
\end{align}
where $\bar{e}_{2z} \equiv e_2(\pi)- e_2(0)$. Equations (\ref{eq:Euler-DIII_2D}) and (\ref{eq:Euler-DIII_3D}) are one of the main results of this paper, which connect the Euler class to the stable topological invariants in class DIII.
In addition, these formulas persist even when $N>2$ by replacing $e_2$ with the second SW class $w_2$,
since $e_2 = w_2 \mod 2$ at $N=2$ and $w_2$ is a stable topological invariant~\cite{Ahn2019stiefel}.

The formulas (\ref{eq:Euler-DIII_2D}) and (\ref{eq:Euler-DIII_3D}) offer a coherent framework for understanding the behavior of topological superconductors under TRSB perturbations that break both $T$ and $C_{2z}$ symmetries but keep their combination intact.
3D DIII superconductors (superfluids) with a nonzero Euler class are dubbed \textit{3D Euler superconductors (superfluids)}.
They remain topologically nontrivial under TRSB perturbations and display robust surface states on an $I_{\rm ST}$ invariant surface.

An example of Euler superfluids is the superfluid $^3$He B phase, which can be modeled as $\mathcal{E} (\bm{k}) = (\bm{k}^2/2m - E_{\rm F}) \mathbbm{1}_2$ and $\Delta(\bm{k}) = \Delta_0 \, \hat{\bm{k}} \cdot \bm{s} \, (is_y)$, 
realizing the 3D topological phase with $ w_{3d} = 1$. 
Here $m$ and $E_{\rm F}$ denote the mass and Fermi energy of $^3$He atoms; 
$\Delta_0$ and $s_i$ represent the amplitude of the pair potential and the Pauli matrices in the spin space.
The number of occupied bands $N=2$ takes the spin degrees of freedom into account.
This system is invariant under $T$ and $C_{2z}$ transformations, where $U_T =is_y $, $U_{C_{2z}} = \exp(i \frac{\pi}{2} \bm{s} \cdot \bm{n}_z)$, and $\bm{n}_z$ is a unit vector whose direction can be arbitrary in the isotropic superfluid but chosen here to be along the $z$ axis.
Thus, the topological phase has a nonzero Euler class, $|\bar{e}_{2z} |=1 \mod 2$.
A typical TRSB perturbation is given by the Zeeman interaction due to magnetic field $\bm{B}$, $\mathcal{E}_{\rm Z} \propto \bm{B} \cdot \bm{s}$, which makes $w_{3d}$ ill-defined.
However, when $\bm{B} \perp \bm{n}_z$, the $C_{2 z}T$ symmetry remains intact, because $T$ reverses $\bm{B}$ and $C_{2z}$ flips it back.
The nontrivial Euler class in a $C_{2z}T$ invariant plane in the 3D BZ ensures the persistence of gapless surface states in the $C_{2z}$ invariant surfaces of the Euler superfluid \cite{suppl}.

The surface states of $^3$He B phase are known to exhibit the Majorana Ising spin response~\cite{Sato2009topo,Chung2009det,Nagato2009,Shindou2010,Mizushima2012sym} and the Majorana hinge states~\cite{volovik2010top,Langbehn2017ref,Geier2018} under magnetic field, and both phenomena stem from the robustness of surface states on the surfaces parallel to the magnetic field. Thus, the existence of a nontrivial Euler class aligns with these earlier findings and offers a novel perspective on the topological properties of $^3$He B.
In addition, the nontrivial second SW class, coming from the relation $e_2 = w_2 \mod 2$, ensures the stability of hinge states~\cite{Ahn2019failure,Ahn2019}.

\begin{table}[]
   \caption{
   \
   Summary of all topological invariants. The first and second columns indicate the Altland-Zirnbauer (AZ) class and the assumed symmetry. The third and fourth columns list possible invariants and provide references to the corresponding formulas.
   }
    \label{tab:formula}
    \centering
    \begin{tabular}{cccc}
    \hline \hline
        AZ  & Symmetry & Invariants &Formulas \\ \hline
        DIII &  $C,T,C_{2z}$ & $\nu_{\rm 2d},\nu_{\rm 3d}, w_{\rm 3d}, e_2,\bar{e}_{2z}$ &  (\ref{eq:DIII_3D}), (\ref{eq:Euler-DIII_2D}), (\ref{eq:Euler-DIII_3D}) \\
        CI  &   $C,T,C_{4z}$ & $\nu_{\rm 4},\bar{\nu}_{\rm 4}, w_{\rm 3d}, e_2,\bar{e}_{2z}$ &  (\ref{eq:CI_3D}), (\ref{eq:Euler-CI_2D}), (\ref{eq:Euler-CI_3D}) \\
        CI  &   $C,T,C_{4z},P$ & $\nu_{\rm 4},\bar{\nu}_{\rm 4}, e_2,\bar{e}_{2z}$ &  (\ref{eq:Euler-CI_2D}), (\ref{eq:Euler-CI_3Dv2}) \\
      \hline \hline   
    \end{tabular}
\end{table}

\textit{Topology of class CI systems.}
We now turn to class CI systems, which are spin-singlet superconductors. This symmetry class supports a stable topological invariant only in 3D space, $w_{\rm 3d} \in 2 \mathbb{Z}$~\cite{Schnyder08}. 
For this symmetry class, $C_{2z}T$ and $PT$ can be used to define $I_{\rm ST}$.

On the one hand, $C_{2z}T$-symmetric and noncentrosymmetric superconductors can have a nontrivial Euler class defined on $C_{2z}T$-invariant planes as in class DIII, which can be related to the 3D winding number characterizing 3D topological superconductors of class CI.
On the other hand, when inversion symmetry is present, $PT$ imposes the real-gauge condition at every $\bm{k}$, enabling to define the Euler class on any 2D closed surface in the 3D BZ.
Since such a closed surface with a nontrivial Euler class cannot be smoothly contracted to a topologically trivial point, a singularity or a node where the excitation gap closes must exist inside the 2D surface~\cite{Bzdusek2017robust}.

We first consider fully gapped CI superconductors or superfluids without inversion symmetry and establish the relationship between $e_2$ and $w_{\rm 3d}$.
To employ the same procedure as in class DIII, we consider $C_{4z}$-symmetric systems.
In order to define a $\mathbb{Z}_2$ index from $C_{4z}$ symmetry, we assume that the relevant energy bands consist of orbital doublets such as ($p_x,p_y$) or $(d_{xz},d_{yz})$~\cite{kobayashi2024}.
Under the $C_{4z}$ operation, these orbitals transform like spatial coordinates ($x,y$), leading to the condition $(U_{C_{4z}})^2 = -\mathbbm{1}_2$. Combining it with $T$ yields the antiunitary operator $C_{4z}T$, which satisfies $(C_{4z}T)^2=-1$. This enforces a Kramers-like degeneracy of the orbital doublets at $C_{4z}T$-invariant momenta: $\Gamma$ and M in the $k_z=0$ plane, and $\Gamma$, M, Z and A in the 3D BZ [see Fig.~\ref{fig:trim} (a)].
For the $k_z=0$ plane, a $\mathbb{Z}_2$ invariant is defined by
\begin{align}
    (-1)^{\nu_{4} } =&
   \frac{\pf[Q(\mathrm{M})U_{C_{4z}}U_T] }{\pf[Q(\Gamma)U_{C_{4z}}U_T]}
   \exp \!\left( \int_\mathrm{M}^{\Gamma} \! \frac{dk}{2}
   \partial_k \ln \{ \det [Q(\bm{k})]\} \! \right) \! ,
 \label{eq:CI_1D}
\end{align}
where $Q(\mathrm{M})U_{C_{4z}}U_T$ and $Q(\Gamma)U_{C_{4z}}U_T$ are skew-symmetric matrices. In the 3D BZ, we can define two $\mathbb{Z}_2$ invariants $\nu_{4} (k_z=0)$ and $\nu_{4}(k_z=\pi)$ involving the integral along the $\Gamma$M and ZA lines, respectively.
Then, the 3D $\mathbb{Z}_2$ invariant is defined as $\bar{\nu}_{4} \equiv \nu_{4}(\pi)- \nu_{4}(0) \mod 2 $. In the absence of inversion symmetry, it is related to the 3D winding number (\ref{eq:w3d}) by~\cite{kobayashi2024}
\begin{align}
    e^{i\frac{\pi}{2}w_{\rm 3d}} = (-1)^{\bar{\nu}_4} . \label{eq:CI_3D}
\end{align}
Note that $w_{3d}\in 2\mathbb{Z}$ in class CI.

Following the same procedure as in class DIII, Eq.~(\ref{eq:CI_1D}) is recast into the following form
\begin{align}
    (-1)^{\nu_{4}} = \frac{\pf [\mathcal{B}_{C_{4z}} (\mathrm{M})]}{\pf[\mathcal{B}_{C_{4z}} (\Gamma)]}. \label{eq:CI_1Dv2}
\end{align}
Therefore, by using Eqs.~(\ref{eq:c4z_e2}), (\ref{eq:CI_3D}), and (\ref{eq:CI_1Dv2}), the relation to the Euler class is given by 
\begin{align}
    \cos \left[ \frac{\pi}{2} e_2 (\bar{k}_z) \right]
     = -\xi_{C_{2z}}(\mathrm{X}) (-1)^{\nu_{4}(\bar{k}_z)}, \label{eq:Euler-CI_2D}
\end{align}
for $\bar{k}_z=0,\pi$, and 
\begin{align}
    \cos \left( \frac{\pi}{2} \bar{e}_{2z}  \right)
    =  e^{i\frac{\pi}{2}w_{\rm 3d}}, \label{eq:Euler-CI_3D}
\end{align}
where the factor $\xi_{C_{2z}}(\mathrm{X}) \xi_{C_{2z}}(\mathrm{R})$ vanishes since the $C_{2z}$ invariant points on the $k_z=0$ and $\pi$ planes share the same $C_{2z}$ eigenvalues unless the gap closes. Since $w_{\rm 3d} \in 2\mathbb{Z}$, the associated $\bar{e}_{2z} $ is also an even integer. 
Thus, Eq.\ (\ref{eq:Euler-CI_3D}) is equivalent to
\begin{equation}
(-1)^{\bar{e}_{2z}/2}=(-1)^{w_{\rm 3d}/2}.
\label{eq:Euler-CI_3D simpler}
\end{equation}
We expect that Eq.\ (\ref{eq:Euler-CI_3D simpler}) can also hold true in $C_{2z}$-symmetric gapped superconductors of class CI, while the relation was derived here assuming $C_{4z}$ symmetry.

However, in the presence of inversion symmetry, Eq.~(\ref{eq:Euler-CI_3D simpler}) no longer holds, while Eq.~(\ref{eq:Euler-CI_2D}) remains valid
and is extended to 
\begin{align}
    (-1)^{\bar{e}_{2z}/2}
    =  (-1)^{\bar{\nu}_{4}}. \label{eq:Euler-CI_3Dv2}
\end{align}
in the 3D BZ, as long as the quasiparticle energy spectra on the $k_z=0$ and $\pi$ planes are fully gapped.
The relation indicates that the centrosymmetric CI superconductors with $\bar{\nu}_4 = 1$ are in a nodal (gapless) phase. This is because a half of the 3D BZ surrounded by $k_z=0$ and $\pi$ planes can be continuously deformed to a sphere [see Fig.~\ref{fig:trim} (b)] while keeping the Euler classes $\bar{e}_{2z}$ nontrivial, implying that a superconducting gap node characterized by the Euler class should be present inside the sphere.
Note that we can define a similar $\mathbb{Z}_2$ invariant in six-fold rotation symmetric systems \cite{kobayashi2024}, to which the approach can be extended \cite{sato2025euler}.

\begin{figure}[tbp]
  \begin{center}
    \includegraphics[width=\linewidth]{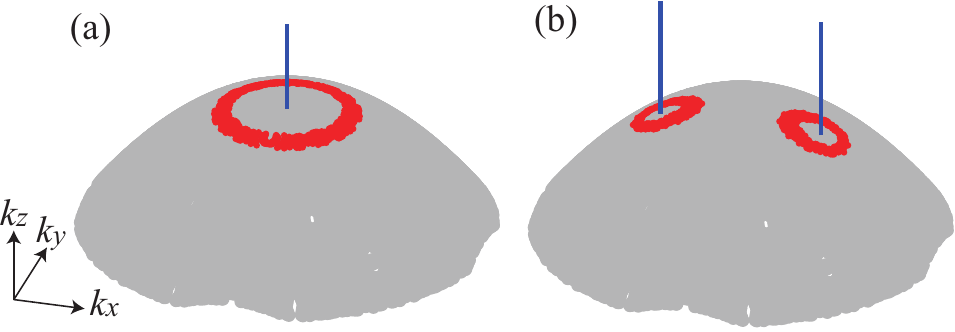}
    \caption{
    (Color online) The inner Fermi surface (gray), superconducting gap nodes (red), and the lines of band degeneracies of the normal-state Hamiltonian (blue) for the BdG Hamiltonian in Eqs.\ (\ref{eq:normal_CI}) and (\ref{eq:gap_CI}) are shown. We change the parameter from (a) $v_4=0$ to (b) $v_4 = 0.0065$ to demonstrate the splitting of the nodal loops. The other parameters are fixed as $(t,\mu,v_1,v_2,v_3,\Delta_0,\Delta_1)=(1,2,0.3,0.3,0.3,0.2,-0.9998)$. 
    }
    \label{fig:nodsal-ring}
  \end{center}
\end{figure}   

To demonstrate this, we consider a tight-binding model of multi-orbital $s_{\pm}$-wave superconductors from ($p_x,p_y$) or ($d_{xz},d_{yz}$) orbital pair, whose Hamiltonian is given by 
\begin{subequations}
\label{eq:3d_model}
\begin{align}
\mathcal{E}(\bm{k}) = \,&
 \left\{ t[\cos(k_x)+\cos(k_y)+\cos(k_z)] -\mu \right\} \mathbbm{1}_2 \nonumber \\
& + v_3 \sin(k_z) \sigma_y 
+ v_1 \sin(k_x) \sin(k_y) \sigma_z \nonumber \\
& + \{ v_2[\cos(k_x) -\cos(k_y)] +v_4\} \sigma_x , \label{eq:normal_CI} \\
\Delta(\bm{k}) =\, & 
\{\Delta_0 + \Delta_1[\cos(k_x)+\cos(k_y)+\cos(k_z)]\}
\mathbbm{1}_2, \label{eq:gap_CI}
\end{align}
\end{subequations}
where  $\sigma_i$ are the Pauli matrices acting in the orbital space, and $\mu$, $t$, $v_i$ are the chemical potential and hopping parameters of two-orbital systems.
We consider an extended $s$-wave pairing parametrized by $\Delta_0$ and $\Delta_1$. Here, the $v_3$ term breaks the inversion symmetry, and the $v_4$ term breaks $C_{4z}$ down to $C_{2z}$.

When $v_3 \neq 0$ and $v_4=0$,  Eq.~(\ref{eq:3d_model}) is invariant under $T$ and $C_{4z}$ transformations ($U_T=\mathbbm{1}_2$ and $U_{C_{4z}}=i \sigma_y$).
From the Fermi surface criterion discussed in Ref.~\cite{kobayashi2024}, we find that the model has $\bar{\nu}_{4}=1$ when we choose the parameters such that the pair potential is nonvanishing and has different signs on the outer and inner Fermi surfaces coming from the orbital doublet that enclose the $\Gamma$ point. 
In this setting, the model hosts $|w_{\rm 3d}| = 2 $, displaying a quadratically dispersing gapless mode on the (001) surface~\cite{kobayashi2024}. The relations (\ref{eq:Euler-CI_2D}) and (\ref{eq:Euler-CI_3D}) imply that the model hosts a nontrivial Euler class $|\bar{e}_{2z} |=2 \mod 4$, and predict that the topological phase is stable under $C_{2z}T$ symmetry-preserving perturbations [see Fig.~\ref{fig:surface} (c)].
  
When $v_3 = v_4 = 0$, the system is additionally invariant under the spatial inversion $U_P=\pm \mathbbm{1}_2$.
From Eq.~(\ref{eq:Euler-CI_3Dv2}), $\bar{\nu}_4=1$ predicts the existence of a superconducting gap node characterized by $|\bar{e}_{2z}| = 2 \mod 4$.
Figure~\ref{fig:nodsal-ring} (a) illustrates a superconducting line node (or a nodal loop) on the inner Fermi surface, together with a line of band degeneracies of the normal-state Hamiltonian at $k_x=k_y=0$. This line node is observed to pass through the nodal loop.
Such linking structure of two line nodes is a universal feature of spectral degeneracies characterized by the nontrivial Euler class or the second SW class~\cite{Ahn2018band,Liu2025Correspondence}.
Furthermore, when we add a small $v_4$ perturbation, the nodal loop splits into two small loops, each of which is characterized by $|\bar{e}_{2z}|=1$ due to the Euler charge conservation. The line of normal-band degeneracies also splits into two lines located at $(k_x,k_y) \simeq  (\pm \sqrt{\frac{2 v_4}{v_2}},0)$ when $v_4/v_2 >0$ as shown in Fig.~\ref{fig:nodsal-ring} (b). Thus, the linking structure persists under the deformations.

\textit{Conclusion.}
We studied the topology of the BdG Hamiltonian with $I_{\rm ST}$ symmetry and showed that topological phases in classes DIII and CI admit a topological characterization by the Euler class. By applying the Euler class formulas [Eqs.~(\ref{eq:c2z_e2}) and (\ref{eq:c4z_e2})] under $C_{nz}$ symmetry, we connected the Euler class with the 2D and 3D topological invariants in class DIII systems, as well as with those in class CI systems with orbital doublets. 
We demonstrated our theory with two examples: the superfluid $^3$He B phase in magnetic field and the nodal lines of multi-orbital $s$-wave superconductors in class CI.
These results provide a theoretical framework for exploring Euler band topology in superfluids and superconductors and offer a unified understanding of the robustness of topological phases under $I_{\rm ST}$-preserving TRSB perturbations, including Majorana Ising physics~\cite{Chung2009det,Nagato2009,Shindou2010,Mizushima2012sym,Dumitrescu2014mag,Mizushima14odd,Xiong2017,Kobayashi2019,Yamazaki2019,Kobayashi2021,Yamazaki2021prb,Yamazaki2021jpsj,Kobayashi2024ptep,Yamazaki2024majo} and higher-order topology~\cite{volovik2010top,Langbehn2017ref,Geier2018,Shapourian2018topo,Wang2018weak,Ghorashi2019second,Yan2019higher,Ahn2020higher,Roy2020higher,Fu2021chiral,Roy2021mixed,Ammar2022higher,Roising2024therm,yamazaki2025higher-order}, which can be probed through spin susceptibility \cite{Chung2009det,Nagato2009,Ominato2024dynamical}, scanning tunneling microscopy~\cite{schindler2018higher}, and transport measurements~\cite{Ikegaya2019}. A promising direction for future research is to explore physical consequences of Euler band topology and to identify candidate materials.  For example, UTe$_2$ \cite{Ran2019,aoki2019} and KFe$_2$As$_2$ \cite{Okazaki2012octet,wu2024nodal} are compelling examples as they satisfy the required symmetry conditions.

\textit{Note added}: After completing this manuscript, we became aware of a related independent study on the Euler topology of different pairing channels in graphene-based models \cite{BJYang_2025}.

 This work was supported by JSPS KAKENHI (Grants No.\ JP22K03478, No.\ JP24K00557, and No.\ JP25K07161) and JST CREST (Grant No.\ JPMJCR19T2).

\bibliography{euler_sc}

\end{document}